\documentstyle[12pt]{article}
\bibliography{unsrt}

\begin{document}
\begin{center}
{\bf {\Large {The Cosmological Quark-Hadron Transition and \\
Massive Compact Halo Objects}}}\\
\vskip 0.2in
Shibaji Banerjee$^a$, Abhijit Bhattacharyya$^b$, Sanjay K. Ghosh$^c$, Sibaji 
Raha$^{a,d}$ \\and \\ Bikash Sinha$^{b,e}$\\
\end{center}
\vskip 0.2in
\noindent
$^a$ Physics Department, Bose Institute, 93/1, A.P.C. Road, Calcutta 700 009, 
INDIA\\
$^b$ Variable Energy Cyclotron Centre, 1/AF, Bidhannagar, Calcutta 700 064, 
INDIA\\
$^c$ Theory Group, TRIUMF, 4004 Wesbrook Mall, Vancouver, B.C. V6T 2A3, 
CANADA\\
$^d$ Nuclear Theory Group, Brookhaven National Laboratory, Bldg. 510A, Upton, 
Long Island, New York 11973-5000, USA\\
$^e$ Saha Institute of Nuclear Physics, 1/AF, Bidhannagar, Calcutta 700 064,
INDIA  
\vskip 0.2in
\noindent
{\bf {One of the abiding mysteries in the so-called standard cosmological 
model is the nature of the dark matter. It is universally accepted that 
there is an abundance of matter in the universe which is non-luminous, due to 
their very weak interaction, if at all, with the other forms of matter, 
excepting of course the gravitational attraction. Speculations as to the 
nature of dark matter are numerous, often bordering on exotics, and searches
for such exotic matter is a very active field of astroparticle physics at the
dawn of the new century. Nevertheless, in recent years, there has been 
experimental evidence \cite{alcock,aubourg} for at least one form of dark 
matter - the massive compact halo objects detected through gravitational 
microlensing effects proposed by Paczynski some years ago \cite{pac}. To 
date, no clear consensus as to what these objects, referred to in the 
literature as well as in the following by the acronym MACHO, are made of; 
for a brief discussion of some of the suggestions, see below. In this work, 
we show that they find a natural explanation as leftover relics from the 
{\it putative} first order cosmic quark - hadron phase transition that is 
predicted by the standard model of particle interactions to have occurred 
during the microsecond epoch of the early universe.}}  
\par
Since the first discovery of MACHO only a few years ago, a lot of effort has 
been spent in studying them observationally, as well as theoretically. It is 
beyond the scope of the present work to cite them adequately; see, for 
example, Sutherland \cite{sut}. Based on about 14 Milky way halo MACHOs 
detected in the direction of the Large Magellanic Cloud (we are not addressing
the events found toward the galactic bulge), the most probable mass estimate 
\cite{sut} for MACHOs is in the vicinity of 0.5$M_{\odot}$, substantially 
higher than the fusion threshold of 0.08$M_{\odot}$. Assuming that MACHOs 
are subject to the limit on total baryon number imposed by the Big Bang 
Nucleosyntheis (BBN), there have been suggestions that they could be white 
dwarfs \cite{fields}. It is difficult to reconcile this with the absence of 
sufficient active progenitors of appropriate masses in the galactic halo. On 
the other hand, there have been suggestions \cite{schramm,jedamzik} that they
could be primordial black holes (PBHs) ($\sim$ 1$M_{\odot}$), arising from 
horizon scale fluctuations triggered by pre-existing density fluctuations 
during the cosmic quark - hadron phase transition. 
While this would not violate the BBN limits on baryon number, the Hawking 
radiation from such primordial black holes would interfere with the observed 
$\gamma$ background, which is thought to be reasonably well understood. It is 
thus safe to conclude that the nature of MACHOs continues to be dark, in the 
sense of begging elucidation.
\par
We propose that the MACHOs are not subject to the BBN limit on baryon number,
insofar as they do not participate in the BBN process, just like the PBHs. On 
the other hand, they do not radiate, via the Hawking process or otherwise,
having evolved out of the quark nuggets which could have been formed in the
cosmic quark - hadron phase transition, at a temperature of $\sim$ 100 MeV
during the microsecond era in the history of the early universe. In a 
seminal work in 1984, Witten \cite{wit} argued that strange quark matter
could be the {\it true} ground state of {\it Quantum Chromodynamics} (QCD),
the underlying field theory of strong interactions and that in a first 
order phase transition from quark - gluon matter to hadronic matter, a 
substantial amount of baryon number could be trapped in the quark phase 
which could evolve into strange quark nuggets (SQNs) through weak 
interactions. QCD - motivated studies of baryon evaporation from SQN-s have 
established \cite{pijush,sumiyoshi} that primordial SQN-s with baryon 
numbers above $\sim$ 10$^{40 - 42}$ would indeed be cosmologically stable. 
More recently, some of the present authors have shown that without much 
fine tuning, these stable SQNs could provide the entire closure density 
($\Omega \sim$ 1) \cite{apj} and in a subsequent work, some of us have 
calculated the distribution of SQN-s produced in the (first order) cosmic 
QCD transition \cite{abhijit1,abhijit2} for various nucleation models, with 
the result that for a reasonable set of parameters, the distribution is 
rather sharply peaked at values of baryon number ($\sim$ 10$^{42-44}$), 
evidently in the stable sector. It was also seen that there were almost no 
SQNs with baryon number exceeding 10$^{46-47}$, comfortably lower than the 
horizon limit of 10$^{49}$ baryons at that time.
\par
It is therefore most relevant to investigate the fate of these SQNs and 
their implications on the later evolution of the universe. While they have
enormous appetite for neutrons, becoming more and more strongly bound in the
process, the total surface area of these large SQNs is not big enough 
to absorb so many neutrons as to interfere with BBN \cite{madsen}. They
remain in equilibrium upto the neutrino decoupling temperature $T_{\nu} \sim$ 
1 MeV beyond which they freeze out. From then on, they are subject only to 
the gravitational interaction. Unlike the usual baryons which are bound by the 
photon pressure till the recombination era, SQNs become free to collapse at
temperatures below $\sim$ 1 MeV. Let us thus roughly estimate the number of 
SQNs contained within the Jeans length at a temperature of 1 MeV. For our
present purpose, we take the SQNs to have the same common mass of 10$^{44}$
with an abundance of 10$^7$ at T $\sim$ 100 MeV \cite{apj}. 
\par
A measure of the Jeans length for the present purpose may be obtained without 
having recourse to the usual hydrodynamic prescription, just by demanding 
that the total gravitational energy in the Jeans volume should be greater 
than or equal to the pressure energy :
\begin{equation}
G(\frac{4}{3} \pi R_J^3 \rho_r)^2/R_J = v_s^2 \rho_r \frac{4}{3} \pi R_J^3
\end{equation}
where the subscript $r$ to $\rho$ indicates that the universe is still 
radiation dominated and $v_s$ stands for the velocity of sound (=$\frac{1}{
\sqrt{3}}$). We then have : 
\begin{equation}
R_J = \frac{m_{pl}}{\sqrt{{4 \pi \rho_r}}} \sim 1.633 t
\end{equation}
which is just less than the distance to the horizon $d_H (\sim 2t$ in the 
radiation era \cite{kolb}). It thus seems that a general relativistic 
treatment is not strictly required, as was to be somewhat expected at least 
for the SQNs, given their enormous mass. 
\par
The number of SQNs within the horizon as a function of temperature is given 
by  $ N_N = 10^7 \left( \frac{100 MeV}{T}\right )^3$ so that the density 
of SQNs is $n_N = N_N / V_H = N_N / \left( \frac{4}{3} \pi (2t)^3 \right)$. 
One can readily see that the total number of SQNs in $R_J$ at $T$ = 1 MeV 
turns out to be $\sim$ 0.58 X 10$^{12-13}$. If all these SQNs clump into one, 
it would then have a mass of $\sim$ 0.5M$_{\odot}$, making them ideal 
MACHO candidates. 
\par
It is obvious that there can be no further clumping of these already clumped 
SQNs; at subsequent times, the density of such objects would be so low that 
it would be hard to find more than one or two of them within one Jeans radius.
A very crude estimate of the collapse time of all the SQNs within $R_J$ can 
be carried out to ascertain that indeed such a timescale is comparable to the
lifetime of the universe at that temperature.
\par
We conclude that gravitational clumping of the primordial SQNs formed in a
first order cosmic quark - hadron phase transition appears to be a plausible 
explanation for the observed halo MACHOs. Needless to say, the estimates 
presented here should serve only as guidelines and a detailed simulation would
of course be needed before any firm conclusions can be drawn. Whether the 
spatial as well as the size distributions of the SQNs can serve as the 
necesary initial fluctuations need also to be carefully looked into. Such a
study is on our present agenda and we hope to present the results in due 
course. 
\par 
The work of S.B. was supported in part by the Council of Scientific \& 
Industrial Research (CSIR), New Delhi. SR would like to thank the Nuclear 
Theory group at Brrokhaven National Laboratory for their warm hospitality 
where part of this work was carried out.

\end{document}